\documentclass[12pt]{article}

\def\noi{\noindent}
\def\barr{\left(\begin{array}}
\def\earr{\end{array}\right)}
\def\beq#1{\begin{equation}\label{#1}}
\def\eeq{\end{equation}}
\def\ber#1{\begin{eqnarray}\label{#1} \nqq}%   left alignment
\def\eer{\end{eqnarray}}

\newcommand{\bear}[1]{\begin{eqnarray}\label{#1}}
\newcommand{\ear}{\end{eqnarray}}

\catcode`\@=11 \@addtoreset{equation}{section}\catcode`\@=12

\newcommand{\N}{ {\bf N} }
\newcommand{\R}{ {\bf R} }

\newcommand{\sh}{\sinh}
\newcommand{\ch}{\cosh}

\newcommand{\sign}{\mathop{\rm sign}\nolimits}

\newcommand{\eps}{\varepsilon}
\newcommand{\tri}{\triangle}

\newcommand{\p}{\partial}
\newcommand{\nn}{\nonumber}

\begin{document}

\begin{center} \large \bf

COMPOSITE S-BRANE SOLUTIONS \\
ON PRODUCT OF RICCI-FLAT SPACES

\end{center}

\vspace{1.03truecm}

\bigskip

\begin{center}

\normalsize
V.D. Ivashchuk, V.N. Melnikov and A.B. Selivanov

\bigskip

Center for Gravitation and Fundamental Metrology,
VNIIMS, 3-1 M. Ulyanovoy Str., Moscow, 119313, Russia and

Institute of Gravitation and Cosmology,
Peoples' Friendship University of Russia,
6 Miklukho-Maklaya St., Moscow 117198, Russia

\end{center}

\centerline{e-mails: ivas@rgs.phys.msu.su, melnikov@rgs.phys.msu.su,
seliv@rgs.phys.msu.su}

\vspace{15pt}

\begin{center}

Key words: S-branes, cosmological solutions, extra dimensions,\\
acceleration.

\end{center}

\vspace{15pt}

\small\noi

\begin{abstract}

A family of generalized $S$-brane solutions with orthogonal
intersection rules and  $n$ Ricci-flat factor spaces in the
theory with several scalar fields and antisymmetric forms  is
considered.  Two subclasses of solutions with power-law and
exponential behaviour of scale factors are singled out. These
subclasses contain sub-families of solutions with accelerated
expansion of certain factor spaces. The solutions depend on
charge densities of branes, their dimensions and intersections,
dilatonic couplings and the number of dilatonic fields.

\end{abstract}

\vspace{20cm}

\pagebreak

\normalsize

%%%%%%%%%%%%%%%%%%%%%%%%%%%%%%%%%%%%%%%%%%%%%%%%%%%%%%%%%%%%%%%%
\section{Introduction}
%%%%%%%%%%%%%%%%%%%%%%%%%%%%%%%%%%%%%%%%%%%%%%%%%%%%%%%%%%%%%%%%

The recent discovery of the cosmic acceleration
\cite{Riess,Perl} was a starting point
for a big number of publications on
multidimensional cosmology giving some explanations of this phenomenon
using certain multidimensional models \cite{Me}, e.g. those of superstring or
supergravity origin (see, for example \cite{IMSjhep} and
references therein).
These solutions deal with time-dependent scale factors of internal spaces
(for reviews see \cite{IMJ,Sam,IMtop,Ierice}) and contain as a special case
the so-called S-brane solutions \cite{S1}, i.e. space like
analogues of $D$-branes \cite{Polc},
see for example
\cite{S2,S3,S4,S5,Isbr,Ohta,Iohta} and references therein.
For earlier $S$-brane solutions see also \cite{BF,LOW,LMPX}.

In our recent paper \cite{IMSjhep} we have obtained
a family of cosmological solutions
with  $(n+1)$ Ricci-flat spaces in the theory with several scalar
fields and multiple exponential potential when
coupling vectors in exponents obey certain
"orthogonality" relations. In \cite{IMSjhep}
two subclasses of "inflationary-type" solutions with power-law
and exponential behaviour of scale factors were found
and solutions with accelerated expansion  were singled out.
In this paper we generalize "inflationary-type"
solutions  from  \cite{IMSjhep} to
$S$-brane configurations in models with
antisymmetric forms and scalar fields. Two subclasses of these
solutions with the power-law and
exponential  behaviour of scale factors in the synchronous time are
singled out.  These subclasses contain sub-families of solutions with
accelerated expansion of certain factor spaces.

Here we deal with a model governed by the action
  \beq{1.1}
   S_g=\int d^Dx
   \sqrt{|g|}\biggl\{R[g]-h_{\alpha\beta}g^{MN}\p_M\varphi^\alpha
   \p_N\varphi^\beta-\sum_{a\in\tri}\frac{\theta_a}{n_a!}
   \exp[2\lambda_a(\varphi)](F^a)^2\biggr\}
  \eeq
where $g=g_{MN}(x)dx^M\otimes dx^N$ is a metric,
$\varphi=(\varphi^\alpha)\in {\R}^{\ l}$ is a vector of scalar
fields, $(h_{\alpha\beta})$ is a  constant symmetric
non-degenerate $l\times l$ matrix $(l\in \N)$, $\theta_a=\pm1$,
%\beq{1.2a}
$F^a =    dA^a
=  \frac{1}{n_a!} F^a_{M_1 \ldots M_{n_a}}
dz^{M_1} \wedge \ldots \wedge dz^{M_{n_a}}$
%\eeq
is a $n_a$-form ($n_a\ge1$), $\lambda_a$ is a
1-form on $\R^l$: $\lambda_a(\varphi)=\lambda_{\alpha a}\varphi^\alpha$,
$a\in\tri$, $\alpha=1,\dots,l$.
In (\ref{1.1})
we denote $|g| =   |\det (g_{MN})|$,
%\beq{1.3a}
$(F^a)^2_g  =
F^a_{M_1 \ldots M_{n_a}} F^a_{N_1 \ldots N_{n_a}}
g^{M_1 N_1} \ldots g^{M_{n_a} N_{n_a}}$,
%\eeq
$a \in \tri$.
Here $\tri$ is some finite set.
For pseudo-Euclidean metric of signature $(-,+, \ldots,+)$
all $\theta_a = 1$.

The paper is organized as following.
In Section 2 we consider
cosmological-type solutions with composite
intersecting $S$-branes from \cite{Isbr,Ierice,Iohta}
on product of Ricci-flat spaces
obeying the "orthogonal" intersection rules.
Section 3 is devoted to exceptional
("inflationary-type") $S$-brane solutions.

%%%%%%%%%%%%%%%%%%%%%%%%%%%%%%%%%%%%%%%%%%%%%%%%%%%%%%%%%%%%%%%%

\section{Cosmological-type solutions with composite
intersecting $p$-branes}
%%%%%%%%%%%%%%%%%%%%%%%%%%%%%%%%%%%%%%%%%%%%%%%%%%%%%%%%%%%%%%%%

\subsection{Solutions with $n$ Ricci-flat spaces}

Let us consider a family of
solutions to field equations corresponding to the action
(\ref{1.1}) and depending upon one variable $u$
\cite{Isbr} (see also \cite{IK,IMtop}).

These solutions are defined on the manifold
  \beq{1.2}
  M =    (u_{-}, u_{+})  \times
  M_1  \times M_2 \times  \ldots \times M_{n},
  \eeq
where $(u_{-}, u_{+})$  is  an interval belonging to $\R$,
and have the following form
\bear{1.3}
  g= \biggl(\prod_{s \in S} [f_s(u)]^{2 d(I_s) h_s/(D-2)} \biggr)
  \biggr\{ \exp(2c^0 u + 2 \bar c^0) w du \otimes du  + \\ \nn
  \sum_{i = 1}^{n} \Bigl(\prod_{s\in S}
  [f_s(u)]^{- 2 h_s  \delta_{i I_s} } \Bigr)
  \exp(2c^i u+ 2 \bar c^i) g^i \biggr\}, \\ \label{1.4}
  \exp(\varphi^\alpha) =
  \left( \prod_{s\in S} f_s^{h_s \chi_s \lambda_{a_s}^\alpha} \right)
  \exp(c^\alpha u + \bar c^\alpha), \\ \label{1.5}
  F^a= \sum_{s \in S} \delta^a_{a_s} {\cal F}^{s},
\ear
$\alpha=1,\dots,l$; $a \in \tri$.

In  (\ref{1.3})  $w = \pm 1$,
$g^i=g_{m_i n_i}^i(y_i) dy_i^{m_i}\otimes dy_i^{n_i}$
is a Ricci-flat  metric on $M_{i}$, $i=  1,\ldots,n$,
  \beq{1.11}
   \delta_{iI}=  \sum_{j\in I} \delta_{ij}
  \eeq
is the indicator of $i$ belonging
to $I$: $\delta_{iI}=  1$ for $i\in I$ and $\delta_{iI}=  0$ otherwise.

The  $p$-brane  set  $S$ is by definition
  \beq{1.6}
  S=  S_e \sqcup S_m, \quad
  S_v=  \sqcup_{a\in\tri}\{a\}\times\{v\}\times\Omega_{a,v},
  \eeq
$v=  e,m$ and $\Omega_{a,e}, \Omega_{a,m} \subset \Omega$,
where $\Omega =   \Omega(n)$  is the set of all non-empty
subsets of $\{ 1, \ldots,n \}$.
Here and in what follows $\sqcup$ means the union
of non-intersecting sets. Any $p$-brane index $s \in S$ has the form
  \beq{1.7}
   s =  (a_s,v_s, I_s),
  \eeq
where
$a_s \in \tri$ is colour index, $v_s =  e,m$ is electro-magnetic
index and the set $I_s \in \Omega_{a_s,v_s}$ describes
the location of $p$-brane worldvolume.

The sets $S_e$ and $S_m$ define electric and magnetic
$p$-branes, correspondingly. In (\ref{1.4})
  \beq{1.8}
   \chi_s  =  +1, -1
  \eeq
for $s \in S_e, S_m$, respectively.
In (\ref{1.5})  forms
  \beq{1.9}
  {\cal F}^s= Q_s  f_{s}^{- 2} du \wedge\tau(I_s),
  \eeq
$s\in S_e$, correspond to electric $p$-branes and
forms
  \beq{1.10}
  {\cal F}^s= Q_s \tau(\bar I_s),
  \eeq
  $s \in S_m$,
correspond to magnetic $p$-branes; $Q_s \neq 0$, $s \in S$.
Here  and in what follows
  \beq{1.13a}
  \bar I \equiv I_0 \setminus I, \qquad I_0 = \{1,\ldots,n \}.
  \eeq

All manifolds $M_{i}$ are assumed to be oriented and
connected and  the volume $d_i$-forms
  \beq{1.12}
  \tau_i  \equiv \sqrt{|g^i(y_i)|}
  \ dy_i^{1} \wedge \ldots \wedge dy_i^{d_i},
  \eeq
and parameters
  \beq{1.12a}
   \varepsilon(i)  \equiv {\rm sign}( \det (g^i_{m_i n_i})) = \pm 1
  \eeq
are well-defined for all $i=  1,\ldots,n$.
Here $d_{i} =   {\rm dim} M_{i}$, $i =   1, \ldots, n$;
$D =   1 + \sum_{i =   1}^{n} d_{i}$. For any
set $I =   \{ i_1, \ldots, i_k \} \in \Omega$, $i_1 < \ldots < i_k$,
we denote
  \bear{1.13}
  \tau(I) \equiv \tau_{i_1}  \wedge \ldots \wedge \tau_{i_k},
  \\
  \label{1.15}
  d(I) \equiv   \sum_{i \in I} d_i, \\
  \label{1.15a}
  \varepsilon(I) \equiv \varepsilon(i_1) \ldots \varepsilon(i_k).
\ear

% $M(I_s)$ is isomorphic to $p$-brane worldvolume manifold
%(see (\ref{1.7})).

The parameters  $h_s$ appearing in the solution
satisfy the relations
\beq{1.16}
  h_s = (B_{s s})^{-1},
\eeq
where
\beq{1.17}
  B_{ss'} \equiv
   d(I_s\cap I_{s'})+\frac{d(I_s)d(I_{s'})}{2-D}+
  \chi_s\chi_{s'}\lambda_{\alpha a_s}\lambda_{\beta a_{s'}}
  h^{\alpha\beta},
\eeq
$s, s' \in S$, with $(h^{\alpha\beta})=(h_{\alpha\beta})^{-1}$.

Here we assume that
\beq{1.17a}
({\bf i}) \qquad B_{ss} \neq 0,
\eeq
for all $s \in S$, and
\beq{1.18b}
({\bf ii}) \qquad B_{s s'} = 0,
\eeq
for $s \neq s'$, i.e. canonical (orthogonal) intersection rules
are satisfied.

The moduli functions read
\bear{1.4.5}
  f_s(u)=
  R_s \sh(\sqrt{C_s}(u-u_s)), \;
  C_s>0, \; h_s \eps_s<0; \\ \label{1.4.7}
  R_s \sin(\sqrt{|C_s|}(u-u_s)), \;
  C_s<0, \; h_s\eps_s<0; \\ \label{1.4.8}
  R_s \ch(\sqrt{C_s}(u-u_s)), \;
  C_s>0, \; h_s\eps_s>0; \\ \label{1.4.9}
  |Q^s||h_s|^{-1/2}(u-u_s), \; C_s=0, \; h_s\eps_s<0,
  \ear
where $R_s = |Q_s|| h_s C_s|^{-1/2}$,
$C_s$, $u_s$  are constants, $s \in S$.

Here
  \beq{1.22}
   \eps_s=(-\eps[g])^{(1-\chi_s)/2}\eps(I_s) \theta_{a_s},
  \eeq
$s\in S$, $\eps[g]\equiv\sign(\det(g_{MN}))$. More explicitly
(\ref{1.22}) reads: $\eps_s=\eps(I_s) \theta_{a_s}$ for
$v_s = e$ and $\eps_s=-\eps[g] \eps(I_s) \theta_{a_s}$  for
$v_s = m$.

Vectors $c=(c^A)= (c^i, c^\alpha)$ and
$\bar c=(\bar c^A)$ obey the following constraints
\beq{1.27}
  \sum_{i \in I_s}d_ic^i-\chi_s\lambda_{a_s\alpha}c^\alpha=0,
  \qquad
  \sum_{i\in I_s}d_i\bar c^i-
  \chi_s\lambda_{a_s\alpha}\bar c^\alpha=0, \quad s \in S,
   \eeq
  \bear{1.30aa}
  c^0 = \sum_{j=1}^n d_j c^j,
  \qquad
  \bar  c^0 = \sum_{j=1}^n d_j \bar c^j,
  \\  \label{1.30a}
  \sum_{s \in S} C_s  h_s +
    h_{\alpha\beta}c^\alpha c^\beta+ \sum_{i=1}^n d_i(c^i)^2
  - \left(\sum_{i=1}^nd_ic^i\right)^2 = 0.
\ear

Here we identify notations  for $g^{i}$  and  $\hat{g}^{i}$, where
$\hat{g}^{i} = p_{i}^{*} g^{i}$ is the
pullback of the metric $g^{i}$  to the manifold  $M$ by the
canonical projection: $p_{i} : M \rightarrow  M_{i}$, $i = 1,
\ldots, n$. An analogous agreement will be also kept for volume forms etc.

Due to (\ref{1.9}) and  (\ref{1.10}), the dimension of
$p$-brane worldvolume $d(I_s)$ is defined by
\beq{1.16a}
  d(I_s)=  n_{a_s}-1, \quad d(I_s)=   D- n_{a_s} -1,
\eeq
for $s \in S_e, S_m$, respectively.
For a $p$-brane we have $p =   p_s =   d(I_s)-1$.

{\bf Restrictions on $p$-brane configurations.}
The solutions  presented above are valid if two
restrictions on the sets of composite $p$-branes are satisfied \cite{IK}.
These restrictions
guarantee  the block-diagonal form of the  energy-momentum tensor
and the existence of the sigma-model representation (without additional
constraints) \cite{IMC}.

The first restriction reads
\beq{1.3.1a}
  {\bf (R1)} \quad d(I \cap J) \leq d(I)  - 2,
\eeq
for any $I,J \in\Omega_{a,v}$, $a\in\tri$, $v= e,m$
(here $d(I) = d(J)$).

The second restriction is following one
\beq{1.3.1b}
  {\bf (R2)} \quad d(I \cap J) \neq 0,
\eeq
for $I\in\Omega_{a,e}$ and $J\in\Omega_{a,m}$, $a \in \tri$.

\subsection{Minisuperspace-covariant notations}

Here  we consider the minisuperspace covariant relations
from \cite{IMJ,IMC} for the sake of completeness. Let
  \beq{2.1}
  (\bar{G}_{AB})=\barr{cc}
  G_{ij}& 0\\
  0& h_{\alpha\beta}
  \earr,
\qquad
(\bar G^{AB})=\left(\begin{array}{cc}
G^{ij}&0\\
0&h^{\alpha\beta}
\end{array}\right)
  \eeq
be, correspondingly,
a (truncated) target space metric and inverse to it,
where  (see \cite{IMZ})
\beq{2.2}
   G_{ij}= d_i \delta_{ij} - d_i d_j, \qquad
   G^{ij}=\frac{\delta^{ij}}{d_i}+\frac1{2-D},
\eeq
and
\beq{2.3}
   U_A^s c^A =
   \sum_{i \in I_s} d_i c^i - \chi_s \lambda_{a_s \alpha} c^{\alpha},
   \quad
   (U_A^s) =  (d_i \delta_{iI_s}, -\chi_s \lambda_{a_s \alpha}),
\eeq
are co-vectors, $s=(a_s,v_s,I_s) \in S$ and
$(c^A)= (c^i, c^\alpha)$.

The scalar product from \cite{IMC} reads
\beq{2.4}
  (U,U')=\bar G^{AB} U_A U'_B,
\eeq
for $U = (U_A), U' = (U'_A) \in \R^N$, $N = n + l$.

The scalar products  for vectors
$U^s$  were calculated in \cite{IMC}
\beq{2.7}
  (U^s,U^{s'})= B_{s s'},
\eeq
where  $s=(a_s,v_s,I_s)$,
$s'=(a_{s'},v_{s'},I_{s'})$ belong to $S$ and
$B_{s s'}$ are defined in (\ref{1.17}).
Due to relations (\ref{1.18b}) $U^s$-vectors
are orthogonal, i.e.
\beq{2.7a}
(U^s,U^{s'})= 0,
\eeq
for $s \neq s'$.

The linear and quadratic constraints
from (\ref{1.27}) and (\ref{1.30a}),
respectively, read in minisuperspace covariant
form as follows:
\beq{2.8}
   U_A^s c^A = 0, \qquad U_A^s \bar{c}^A = 0,
   \eeq
  $s \in S$,
and
\beq{2.10}
  \sum_{s \in S} C_s  h_s +
  \bar G_{AB} c^A c^B = 0.
\eeq

%%%%%%%%%%%%%%%%%%%%%%%%%%%%%%%%%%%%%%%%%%%%%%%%%%%%%%%%%%%%%%%%%%%%
  \section{Special solutions}
%%%%%%%%%%%%%%%%%%%%%%%%%%%%%%%%%%%%%%%%%%%%%%%%%%%%%%%%%%%%%%%%%%%%

Now we consider a special case of classical solutions
from the previous section when $C_s = u_s = c^i = c^{\alpha} = 0$
and
\beq{6.0}
   B_{ss} \eps_s < 0,
\eeq
$s \in S$.

We get two families of solutions written in synchronous-type
variable with:

  A) power-law dependence of scale factors for $B \neq -1$,

  B) exponential dependence of scale factors for $B  = -1$,

where

\beq{6.1}
  B = \sum_{s \in S} h_s \frac{d(I_s)}{D - 2}.
\eeq

Remind that $h_s= (B_{ss})^{-1}$.

\subsection{Power-law solutions}

Let us consider the solution corresponding to
the case $B   \neq -1$. The solution reads
\bear{6.2}
  g=  w d\tau \otimes d\tau
  +   \sum_{i = 1}^{n} A_i \tau^{2 \nu_i} \hat{g}^i, \\
  \label{6.3}
  \varphi^\alpha= \frac{1}{B+1}
  \sum_{s \in S} \chi_s h_s \lambda_{a_s}^{\alpha} \ln \tau   +
  \varphi^{\alpha}_0,
\ear
where $\tau > 0$,
\beq{6.4}
  \nu_{i} = - \frac{1}{B+1} \sum_{s\in S} h_s
  \left(\delta_{iI_s} - \frac{d(I_s)}{D-2} \right),
\eeq
$i = 1, \dots, n$ and
\beq{6.5}
|h_s| \left( \prod_{i \in \bar{I}_s} A_i^{d_i} \right)
  \exp( 2 \chi_s \lambda_{a_s \alpha} \varphi^{\alpha}_0)
  = Q_s^2 |B+1|^2,
\eeq
$s \in S$; and $A_i > 0$ are arbitrary constants.

The elementary forms read
  \beq{6.9}
  {\cal F}^s=  \frac{|h_{s}| A^{1/2}}{Q_s (B+1)|B+1|}
\tau^{-(B+2)/(B+1)}  d\tau \wedge \tau(I_s),
  \eeq
$s \in S_e $, (for electric case) and
forms
  \beq{6.10}
  {\cal F}^s= Q_s \tau(\bar I_s),
  \eeq
  $s \in S_m$, (for magnetic case). Here
  and in what follows  $w = \pm 1$, $Q_s \neq 0$, $s \in S$, and
  $A = \prod_{i=1}^{n} A_i^{d_i}$.

We see that these
solutions depend on charged densities of branes, their dimensions and
intersections, dilatonic couplings and the number of dilatonic
fields.

In the special case of electric $S$-branes of maximal
dimension $d(I_s) = D - 1$ the metric
and scalar fields are coinciding (up to notations)
with the solutions obtained in \cite{IMSjhep}
when signature restrictions (\ref{6.0}) are obeyed. Since
solutions from \cite{IMSjhep} contain a subfamily
of solutions with accelerated expansion of factor spaces,
we are led to non-empty set of solutions with "acceleration"
in the model under consideration  \cite{IMSacc}.

\subsection{Solutions with exponential scale factors}

Here we consider the solution corresponding to
the case $B = -1$. The solution reads
\bear{6.15}
    g=  w d\tau \otimes d\tau
    +  \sum_{i = 1}^{n} A_i \exp(2 M \mu_i \tau) g^i, \\
    \label{6.16}
    \varphi^{\alpha} = - M \tau \sum_{s\in S}
    h_s \chi_s \lambda_{a_s}^{\alpha}
    + \varphi^{\alpha}_0,
\ear
where
\beq{6.17}
    Q_s^2 \exp(- 2 \chi_s \lambda_{a_s \alpha} \varphi^{\alpha}_0)
    = |h_s| M^2 \prod_{i \in \bar{I}_s} A_i^{d_i},
\eeq
$s \in S$,
\beq{6.18}
  \mu_{i} = \sum_{s \in S} h_s
  \left(\delta_{iI_s} - \frac{d(I_s)}{D-2} \right),
\eeq
$M$ is parameter
and $A_i > 0$ are arbitrary constants, $i = 1, \dots, n$.

The elementary forms read
  \beq{6.19}
  {\cal F}^s=  \frac{|h_{s}| A^{1/2}}{Q_s}
M^2 e^{ M \tau} d\tau \wedge \tau(I_s),
  \eeq
for $s \in S_e$,  and
${\cal F}^s= Q_s \tau(\bar I_s)$ for $s \in S_m$.

In the cosmological case $w = -1$ we get an accelerated
expansion of factor space $M_i$ if and only if $\mu_i M > 0$
\cite{IMSacc}. We see again that these
solutions depend on  charge densities of branes, their dimensions and
intersections, dilatonic couplings and the number of dilatonic
fields.

\section{Conclusions}

In this paper we considered  generalized $S$-brane solutions
with orthogonal intersection rules
and  $n$ Ricci-flat factor spaces in the theory with several scalar
fields and antisymmetric forms. We singled out  subclasses of solutions
with power-law and  exponential behaviour of scale factors
depending in general on charge densities of branes, their dimensions
and intersections, dilatonic couplings and the number of dilatonic fields.
These subclasses contain sub-families of solutions with accelerated
expansion of certain factor spaces \cite{IMSacc}, e.g.  those considered
in our earlier paper \cite{IMSjhep} (with signature restrictions
(\ref{6.0}) imposed).

We note that in our approach the intersection rules
for composite $S$-branes have a minisuperspace covariant form, i.e.
they are formulated in terms of scalar products
of  brane $U$-vectors and generally (see \cite{Isbr}) are classified
by Cartan matricies of (semi-simple) Lie algebras.
The intersection rules considered in this paper
correspond to the Lie algebra $A_1 + \ldots + A_1$.

\begin{center}
{\bf Acknowledgments}
\end{center}

This work was supported in part by the Russian Ministry of
Science and Technology, Russian Foundation for Basic Research
(RFFI-01-02-17312-a) and Project DFG (436 RUS 113/678/0-1(R)).
V.N.M. is grateful to Prof. J.-M. Alimi for the hospitality
during his stay at LUTH, Observatory Paris-Meudon, France.

\end{document}